\documentclass[11pt,twoside]{article}

\usepackage{prcsa2008}
\usepackage{epsf}
\usepackage{psfig}
\usepackage{lscape}

\usepackage{graphicx} 

\begin{document}
\title{Towards a complete census of young stars in the solar neighbourhood with SkyMapper}  
\author{Simon J. Murphy and Michael S. Bessell} 
\affil{Research School of Astronomy and Astrophysics\\ Australian National University\\ Cotter Road, Weston Creek, ACT 2611, Australia}   

\begin{abstract} 
In this contribution we outline plans for identifying and characterising numerous young, low-mass stars within 150 pc of the Sun using the new SkyMapper telescope and Southern Sky Survey. We aim to learn more about the star formation history of the solar neighbourhood over the past 5--50 Myr, the dispersal processes involved, as well as testing pre-main sequence evolutionary models and the universality of the stellar Inital Mass Function. Searching for the dispersed halo of low-mass objects predicted to surround the $\eta$ Chamaeleontis cluster will be one of the first goals of the project.
\end{abstract}

\section*{Introduction}

In the past decade it has been possible to discern numerous coeval, comoving associations of young stars within the solar neighbourhood. These include the TW Hydrae and Tucana-Horologium associations, the young open cluster $\eta$~Chamaeleontis, and the $\beta$ Pictoris and AB Doradus Moving Groups \citep[see][for a recent review]{SM2_Zuckerman04}. These sparse groups of young stars are spread over the entire southern sky and appear kinematically tied to the Scorpius-Centaurus OB association, an extensive region of recent star formation near the Sun \citep[Makarov, these proccedings;][]{SM2_Fernandez08}. 

Populations of young, nearby stars are interesting for a variety of astrophysical reasons. The 5--50~Myr ages of these stars are intermediate between obscured 1--2~Myr old T Tauri stars in denser star forming regions such as Orion and $\sim$100 Myr old open clusters like the Pleiades. This is the same timescale during which gaseous, and later terrestrial, planets are formed from circumstellar disks. By observing stars in these young local associations we are afforded a unique view of disk evolution and planet formation at high angular resolution. Previous large-scale surveys have traditionally been biased toward the brighter, more massive association members. To rectify this we aim to identify and characterise complete samples of the low-mass ($<$$1\textrm{ M}_{\odot}$) populations of these associations. Studies of stellar angular momentum evolution, multiplicity and testing the consistency of pre-main sequence evolutionary models all benefit from statistical samples of nearby, coeval, codistant stars in this age and mass range. Furthermore, a complete census of low-mass members allows the university of the stellar Initial Mass Function in different  environments to be investigated. 

In piecing together the recent star formation history of the solar neighbourhood, exact time-sequences and causality are difficult to constrain without knowledge of the low-mass population. Massive OB stars essentially form on the main sequence and evolve off  again on timescales of several Myr. Low-mass stars however have a more protracted pre-main sequence evolution and are much more numerous, better tracing the fossil remnants of star formation and the dispersal processes involved as young clusters disintegrate into the Galactic field.

\section*{Candidate selection}

Selection of candidate young, low-mass stars will be from photometry and astrometry produced by the SkyMapper telescope and Southern Sky Survey \citep[Murphy et al., these proceedings;][]{SM2_Keller07}. SkyMapper will provide multi-epoch $uvgriz$ photometry over the entire southern sky to a depth of $g=23$ mag with  0.03 mag precision and astrometry to 50 mas. We propose a three-phase observational programme, exploiting the strengths of the 
SkyMapper facility: accurate wide-field photometry, astrometric precision and the ability 
to observe large areas at multiple epochs. 

\subsubsection{Photometry.}
By combining SkyMapper $riz$ photometry with 2MASS $JHK_{s}$ and DENIS $iJK_{s}$ we can robustly select late-type candidates over wide fields. The addition of non-survey H$\alpha$ imaging on the same instrument allows chromospheric activity, as measured through H$\alpha$ emission, to be used as a proxy for age, as well as identifying accretion-driven emission in the most youthful of candidates (see Figure~\ref{SM_fig:halpha}). The first SkyMapper data product will be the Five Second Survey - a three epoch survey in all filters of stars between 8th and 15th magnitude in photometric conditions. Stacked 15 s exposures (3$\times$5 s) allow us to reach a limiting magnitude of $i\sim18.5$ at a signal to noise of 10--20.  At the distance of TWA (55 pc) this is a main sequence spectral type of M8, whereas by $\eta$ Cha (97 pc) it is M6, and by Upper Scorpius (150 pc) it falls again to M5. However, because pre-main sequence stars can lie 
several magnitudes above the Zero Age Main Sequence at ages 10 to 100 Myr, 
this allows us to probe several spectral types later than the limits 
given above. 2MASS $JHK_{s}$ photometry is available for all stars earlier than these 
limits. Near-IR photometry will be especially useful in reducing contamination from 
giant stars, which trace a separate locus to dwarfs on two-colour diagrams. 

\subsubsection{Astrometry.}

Proper motions are a valuable tool for determining association membership. Nearby association members may be spread over many (tens or hundreds of) square degrees on the sky but should possess common proper motions. Combining initial SkyMapper observations with photographic POSS-II plates gives a 20 to 30 year baseline over which to compute proper motions. This should yield individual proper motions accurate to $\sim$7 mas yr$^{-1}$. The full complement of nine SkyMapper epochs over five years improves this accuracy to 4 mas yr$^{-1}$. Members of local associations exhibit proper motions on order of several tens of mas yr$^{-1}$, suggesting our initial epochs and POSS-II data should be adequate for kinematic selection. Three-dimensional space motions are required to kinematically trace associations back to their parent star-forming regions and analyse any encounters with other clusters. Space velocities can be calculated from proper motions with knowledge of the radial velocity (from spectra) and distance (from parallax, evolutionary tracks, moving cluster method etc.). 

\section*{Dynamical evolution of the $\eta$ Chamaeleontis cluster}

A thorough search for the low-mass population of the $\eta$ Chamaeleontis open cluster \citep{SM2_Mamajek99} aptly demonstrates the capabilities of  SkyMapper's wide-field photometry and precision astrometry. This nearby (97~pc) 8~Myr old cluster contains 18 systems and 27 known members with an additional $\sim$30 low-mass and substellar members expected from the cluster Initial Mass Function \citep{SM2_Lyo04}. However, recent searches for this missing population have not identified any additional members at either large radii \citep[][to 4~times the radius of known membership]{SM2_Luhman04} or to low stellar masses in the cluster core \citep[][to 0.013 M$_{\odot}$]{SM2_Lyo06}.  Failure to find these low-mass members raises a fundamental question: is the cluster's evolution driven by dynamical interactions which dispersed the stars into a diffuse halo at large radii or does $\eta$~Cha possess an abnormally top-heavy IMF deficient in low-mass objects?

\begin{figure}[tb!] 
   \centering
   \includegraphics[viewport=0 10 460 420,clip,width=0.61\linewidth]{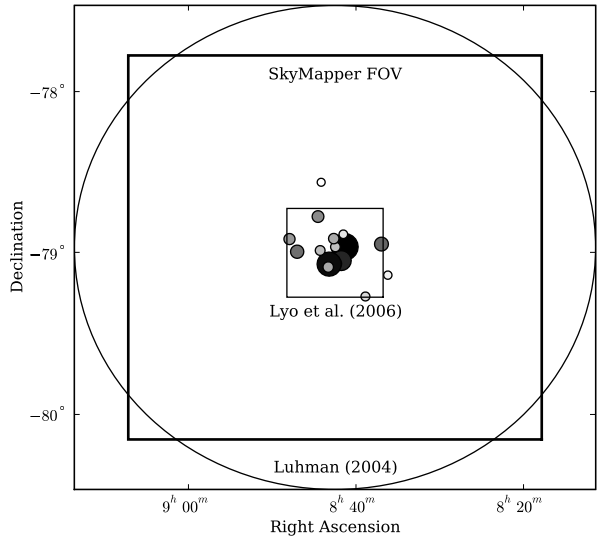} 
   \caption{The 18 member systems of the $\eta$ Cha cluster. Masses of the members range from 0.15 to 3.4 M$_{\odot}$ and are given by the size and shading of the points. Also shown is the 1.5\deg~search radius adopted by \citet{SM2_Luhman04} and the $33'\times33'$ region surveyed to 0.013 M$_{\odot}$ by \citet{SM2_Lyo06}. Shown for comparison is a single 5.7 deg$^{2}$ SkyMapper field centered on the cluster.}
   \label{SM_fig:etacha}
\end{figure}

Figure~\ref{SM_fig:etacha} illustrates the clear advantage of SkyMapper's wide-field photometry over previous membership surveys. A single 5.7 deg$^{2}$ SkyMapper field centered on the cluster easily covers the known members and almost spans the 1.5\deg~radius searched by \citet{SM2_Luhman04}. We propose to search for members of a low-mass halo surrounding $\eta$ Cha
at radii of up to 5\deg~from the cluster core (9~pc at the cluster distance). Recent dynamical modeling by \citet{SM2_Moraux07} suggest that should dynamical evolution have played a dominant role in shaping the present configuration of the cluster, the majority of low-mass objects will have been ejected to radii beyond that currently surveyed. We should expect to find a handful of objects down to 0.1 M$_{\odot}$ in such a search area. Stacked 15~s exposures from the Five Second Survey allow us to reach depths of $i\sim18.5$, similar to the DENIS photometry used by \citet{SM2_Luhman04}. By coadding images from the deeper Main Survey we can easily reach $i\sim20$ with 110 s exposures. For an 8~Myr old population at 97~pc these limits are sufficient to probe down to the brown dwarf-planet boundary near 0.015 M$_{\odot}\simeq$ 15 M$_{Jup}$.

\begin{figure}[tb!] 
   \centering
   \includegraphics[width=0.62\linewidth]{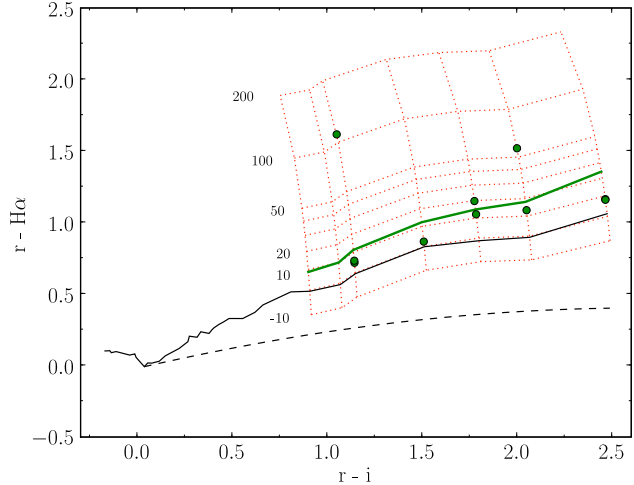} 
   \caption{Synthetic $r-i$ vs. $r-\textrm{H}\alpha$ colours for the M-type members of $\eta$~Cha (closed circles). The grid shows the effect of adding H$\alpha$ emission lines of varying EW (from -10 \AA~absorption to +200 \AA~emission) to M0-M6 library spectra. Spectral types and H$\alpha$ equivalent widths of members are taken from the literature. Field stars are expected to lie between the non-reddened main seqeunce (solid line) and the A0 reddening line (dashed line). The empirical spectral type-EW accretion threshold of \citet{SM2_Barrado03} is also shown.}
   \label{SM_fig:halpha}
\end{figure}

The surface density of candidates is small enough ($\sim$0.05 stars deg$^{-2}$) that it would be difficult to distinguish this dispersed population from photometry alone. By combining photometry with proper motions we will be better able to select ejected cluster members as their projected velocities should be distributed around the mean cluster motion of (-29.9, +27.5) mas yr$^{-1}$. The addition of H$\alpha$ imaging allows a further diagnostic - all known $\eta$~Cha M-dwarfs exhibit chromospheric H$\alpha$ emission due to their youth. Several systems with larger H$\alpha$ emission have been confirmed as actively accreting. Diagrams similar to Figure~\ref{SM_fig:halpha} will be useful for both selecting young, late-type objects as well as estimating photometric equivalent widths. H$\alpha$ emission should manifest itself as an $r-\textrm{H}\alpha$ colour excess above the locus of non-emitting dwarfs. The different molecular band structure in background giants also allows them to be distinguished on the basis of $r-\textrm{H}\alpha$ colour, aiding in reducing contamination.

\end{document}